\documentclass[useAMS]{mn2e}
\usepackage{epsfig,natbib_vw,times}

\bibpunct[, ]{(}{)}{;}{a}{}{,}

\def \aj {AJ}
\def \mnras {MNRAS}
\def \apj {ApJ}

\def \apjl {ApJL}

\def \pasp {PASP}
\def \aaps {AAPS}

\def \be {\begin{equation}}
\def \ee {\end{equation}}
\def \ul {\underline}
\def\gsim{\mathrel{\lower0.6ex\hbox{$\buildrel {\textstyle >}
 \over {\scriptstyle \sim}$}}}
\def\lsim{\mathrel{\lower0.6ex\hbox{$\buildrel {\textstyle <}
 \over {\scriptstyle \sim}$}}}

\def\m@th{\mathsurround=0pt }
\def\eqalign#1{\null\,\vcenter{\openup1\jot \m@th
 \ialign{\strut\hfil$\displaystyle{##}$&$\displaystyle{{}##}$\hfil
 \crcr#1\crcr}}\,}

\def \nskyspec {15,178 \ }

\title[Peering through the OH-forest]{Peering through the OH-forest: a new technique to remove
  residual sky features from SDSS spectra}
\author[V. Wild \& P. Hewett]{Vivienne Wild\thanks{vw@ast.cam.ac.uk},
  Paul C. Hewett
\vspace*{6pt}\\
Institute of Astronomy, University of Cambridge, Madingley Road,
Cambridge CB3 0HA, UK \\}

\begin{document}
\maketitle
\begin{abstract}

The Sloan Digital Sky Survey (SDSS) currently provides by far the largest
homogeneous sample of intermediate signal-to-noise ratio (S/N) optical spectra
of galaxies and quasars.  The fully automated SDSS spectroscopic reduction
pipeline has provided spectra of unprecedented quality that cover the wavelength
range $3800-9200\,$\AA.  However, in common with spectra from virtually all
multi-object surveys employing fibres, there remain significant systematic
residuals in many of the spectra due to the incomplete subtraction of the strong
OH sky emission lines longward of $6700\,$\AA.  These sky lines affect almost half
the wavelength range of the SDSS spectra, and the S/N over substantial wavelength
regions in many spectra is reduced by more than a factor of 2 over that expected
from counting statistics.  We present a method to automatically remove the sky
residual signal, using a principal component analysis (PCA) which takes
advantage of the correlation in the form of the sky subtraction residuals
present in each spectrum.  Application of the method results in spectra with
essentially no evidence for degradation due to the incomplete subtraction of OH
emission features.  A dramatic improvement in the quality of a substantial
number of spectra, particularly those of faint objects such as the bulk of the
high-redshift quasars, is achieved.  We make available IDL code and
documentation to implement the sky residual subtraction scheme on SDSS
spectra included in the public data releases. To ensure that
absorption and emission features intrinsic to the object spectra do not affect
the subtraction procedure line masks must be created that depend on
the scientific application of interest.  We 
illustrate the power of the sky-residual subtraction scheme using samples of
SDSS galaxy and quasar spectra, presenting tests involving the near-infrared
CaII triplet absorption, metal absorption line features in damped Ly$\alpha$
systems and composite spectra of high-redshift quasars.

\end{abstract}


\section{Introduction}

The development over the last two decades of efficient and reliable wide-field
multi-object spectrographs has resulted in enormous advances in the ability to
compile large samples of high-quality astronomical spectra.  The Sloan Digital
Sky Survey \citep[SDSS;][]{2000AJ....120.1579Y} represents the most impressive
application yet of such an instrument to provide spectroscopic samples of
quasars, galaxies and stars of unprecedented size.  The third data release
(DR3; Abazajian et al. 2004b)\nocite{astro-ph/0410239} provides the astronomical community with
nearly 500,000 object spectra and further releases will take the number to
750,000 within eighteen months.

The SDSS spectra present a dramatic improvement in quality over data from
previous surveys:  extended wavelength coverage, $3800-9200\,$\AA; intermediate
resolution, $\lambda/\Delta\lambda \simeq 2500$; high-quality relative and
absolute flux-calibration; typical signal-to-noise ratios (S/N) of $10-30$;
availability of accurate noise and mask arrays.  Coupled with the overall
homogeneity of the dataset this makes them suitable for an almost limitless
number of quantitative investigations.

A particularly impressive aspect of the instrumental design, observing strategy
and data reduction pipeline \citep{astro-ph/0408167} has been the quality of the
sky-subtraction achieved considering the relatively large, 3 arcsecond diameter
fibres and fully automated reduction.  Subsets of the SDSS spectra, such as the
fainter quasars, include objects with magnitudes in red passbands equivalent to
the sky-brightness per square arcsecond.  Despite the demonstrable quality of
the SDSS spectra, visual inspection reveals significant systematic
sky-subtraction residuals longward of $6700\,$\AA \ in many spectra.  For
fainter objects in particular, the sky-subtraction residuals are the dominant
source of uncertainty over a wavelength interval of some $2000\,$\AA.  The
spectroscopic region involved is extensive and includes features of significant
astrophysical interest; examples include the Calcium triplet (8500, 8545,
$8665\,$\AA), a powerful diagnostic of stellar populations in low-redshift galaxies,
and the H$\beta$ + [OIII] 4959, $5007\,$\AA \ emission region in quasars and active
galactic nuclei (AGN) in the redshift interval $0.4 < z < 0.8$.

With sufficient care \citep[e.g.][]{2004AJ....127.1860B} it is possible to
quantify the noise properties of the SDSS spectra in the red, allowing the
statistical significance of features to be estimated reliably.  However, the
decrease in the S/N relative to that expected from counting statistics can be
factors of 2-3.  Coupled with the substantial fraction of the spectral range
affected, many potential scientific investigations are compromised.

The sky-subtraction procedures incorporated in the SDSS spectroscopic pipeline
are very effective, overcoming a number of the longstanding problems associated
with wide-field multi-fibre observations.  Brief details can be found in the
Early Data Release paper \citep[][Sections 4.8.5 and
4.10.1]{2002AJ....123..485S} and information on calibration improvements for the
Second Data Release (DR2) are given in Abazajian et al. (2004a)\nocite{2004AJ....128..502A}.  While the
large area of sky, $7.1\,$ square arcseconds, observed through the fibre
aperture exacerbates the problem of achieving accurate sky-subtraction, the
origin of the dominant systematic residuals present in the SDSS spectra is not a
preserve of fibre spectroscopy alone.  Rather, the fundamental issue lies in the
ability to remove OH emission features from spectra in which the line profiles
are barely resolved.  Combined with sub-pixel changes in the pixel-to-wavelength
calibration between spectra, sharp residuals often remain.

\begin{figure*}
\vspace*{-7mm}
\begin{minipage}{\textwidth}
\hspace*{-1cm}
\includegraphics[scale=1.05]{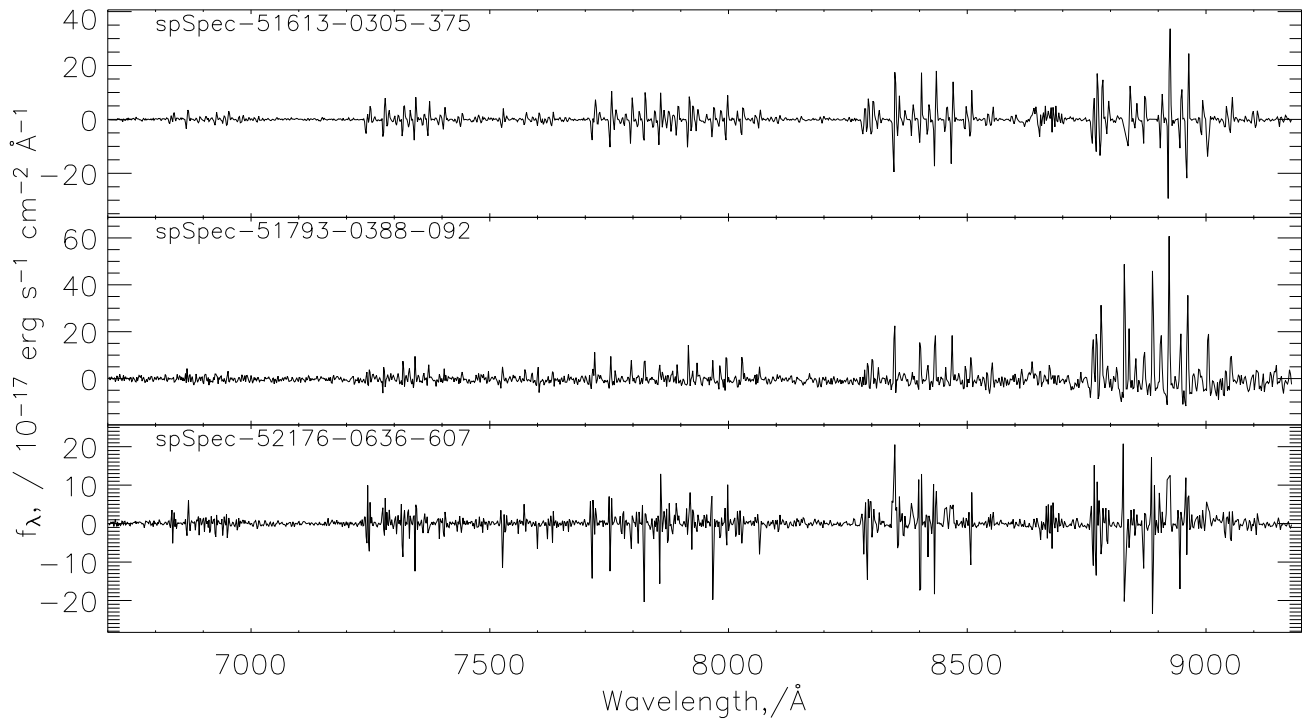}
\vspace*{-7mm}
\caption{Examples of poor OH sky emission line subtraction in SDSS
sky-subtracted sky spectra. Note the well-defined patterns present,
particularly the characteristic positive-negative residuals.}
\label{fig_skyegs}
\end{minipage}
\end{figure*}

\citet{2003PASP..115..688K} provides a recent summary of the difficulties
associated with accurate sky-subtraction in long-slit observations, presenting a
highly effective procedure for removing the signature of even rapidly varying
and poorly sampled emission line features.  The technique described by Kelson is
certainly not applicable to the extracted fibre spectra available as part of the
official SDSS Data Releases and any improvement must rely on an alternative
procedure.  The sky-subtraction residuals arise from the subtraction of two
essentially identical tooth-comb signatures that have been very slightly
misaligned relative to one another leading to well-defined patterns (Fig.
\ref{fig_skyegs}).  This suggests a correction procedure that takes advantage of
the correlations present in the wavelength direction, offering significant
advantages over simply masking the affected pixels.

More specifically, we develop an approach for the removal of the dominant OH
sky-subtraction residuals in the SDSS spectra based on principal component
analysis (PCA).  PCA (also called Karhunen--Lo\`{e}ve transformation)
is a well-established data reduction technique in astronomy, frequently applied to
classification and reconstruction problems associated with large numbers of
spectra of various types of object (e.g.  galaxies:
\nocite{1995AJ....110.1071C,1996MNRAS.283..651F,2002MNRAS.333..133M,
2003ApJ...599..997M,2004AJ....128..585Y,1992ApJ...398..476F,astro-ph/0408578,
1983A&AS...51..443W} Connolly et al. 1995; Folkes et al. 1996;
Madgwick et al. 2002, 2003; Yip et al. 2004a.  Quasars:  Francis et
al. 1992; Yip et al. 2004b.  Stars:  Whitney 1983).  The
idea of employing PCA for sky-subtraction has been suggested by
\citet{2000ApJ...533L.183K}, who present a complex method to remove the sky
signal without the need for concurrent sky observations.  However, the scheme
appears to have generated little interest, perhaps because of the ambitious goal
of the technique and the targeting of observations in which sky spectra are
unavailable.  By contrast, we develop a much more specific application of the
PCA technique to the SDSS DR2 spectra that results in an improvement by over a
factor of 2 in the S/N of those pixels longward of $6700\,$\AA\, affected by OH
sky lines, dramatically increasing the potential use of the red half of the SDSS
spectra for a variety of scientific investigations.  The technique is equally
applicable to the spectra in the recent DR3 release. We adopt the same 
convention as employed in the SDSS and use vacuum wavelengths throughout the
paper.

In Section \ref{sec_method} we present our method which is applied to SDSS
galaxy and quasar spectra in Section \ref{sec_recon}.  Section \ref{sec_tests}
uses subsets of the SDSS DR2 spectroscopic catalogue to demonstrate application
of the method to various astronomical studies. The code and
documentation with which to carry out the method on public release SDSS spectra is
available on the web at {\tt http://www.ast.cam.ac.uk/research/downloads\\/code/vw/}.

\section{Method} \label{sec_method}

Our overall approach is determined by the extent of the spectroscopic
information provided in a SDSS public data release. The raw CCD exposures,
typically 4 or 5 per spectroscopic plate, that make up the total exposure in each 
spectroscopic field are not available, precluding any attempt to improve
the sky-subtraction on an exposure by exposure basis.
However, the SDSS data releases do provide the final reduced spectra for both 
target objects and for the fibres allocated to blank sky regions employed to 
define the underlying spectrum of the night sky. These contain 
significant diagnostic information about the quality of sky subtraction for each 
spectroscopic plate.

For each observation of a SDSS spectroscopic plate, approximately 32
fibres, 16 for each of
the 2 spectrographs, are assigned to blank sky regions, selected from areas
containing no detected objects in the SDSS imaging survey.  The sky fibres are
identified as ``{\small SKY}'' in the catalogue's ``{\small OBJTYPE}'' and
``{\small SPECCLASS}'' field.  The sky spectra are combined to create a
``master-sky'' spectrum for each spectroscopic plate, which is then scaled and
subtracted from each of the 640 spectra, producing 608 sky-subtracted object
spectra 
and 32 sky-subtracted sky spectra, all of which are part of the
standard SDSS data releases. Throughout the remainder of the
paper we will refer to the sky-subtracted sky spectra as ``sky
spectra'', and the sky-subtracted object spectra as ``object spectra''.

The availability of large numbers of sky spectra,
over 18,500 in DR2, thus provides a direct empirical measure of the
noise resulting from counting statistics as well as any systematic residuals
from the sky-subtraction procedure.  The scale of the SDSS releases is such that
the application of empirical self-calibration techniques, based on the
statistical properties of the data set itself, can be a powerful tool.

Visual inspection of the spectra of faint objects, or sky
spectra (Fig.  \ref{fig_skyegs}), immediately reveals the presence of
``patterns'' due to small imperfections in sky subtraction.  The presence of
systematic deviations that are correlated over extended wavelength ranges
suggests that a technique capable of quantifying the form and amplitude of the
correlated deviations could allow the removal of a substantial fraction of the
sky-subtraction noise.  PCA is a well-established
technique with a number of desirable properties for such an application.

The principles behind our sky-subtraction method are simple to understand:  a
PCA of the sky spectra produces a set of orthogonal components
that provides a compact representation of the systematic residuals
resulting from the sky-subtraction process.  The components are then added in
linear combinations to remove the systematic sky residuals from the spectra of
target objects, such as galaxies.

Our analysis focuses on the red part of the SDSS spectra ($6700-9180\,$\AA), as 
the strong emission features blueward of $6700\,$\AA \  are small in number and
easily masked. The upper limit of $9180\,$\AA \ is set by the requirement that the
entire wavelength region be included in all spectra when creating the PCA
components from the sky spectra. Although techniques exist to overcome
this \citep{1999AJ....117.2052C}, it was believed they would not greatly
improve the applicability of the procedure for scientific objectives
due to the very small fraction of plates affected.

\subsection{A sample of sky spectra}

The goal of the sky-subtraction scheme is to remove any systematic residuals
from the bulk of the SDSS spectra.  Pathological spectra with very unusual
deviations can have a disproportionate effect on the generation of the PCA
components.  Therefore, we identify a restricted subset of the full 18,764 sky
spectra for use in deriving the PCA components.

For the wavelength range used in the analysis ($6700-9180\,$\AA)
the spectra must satisfy the following criteria, where the numerical
values are in the units of the SDSS spectra ($10^{-17} {\rm erg \ s^{-1} 
cm^{-2} \AA^{-1}}$):
\begin{enumerate}
\item $-0.2<$ mean flux $<0.2$, ensuring the spectra have a mean close to zero.
\item variance of the flux $<0.8$, ensuring that the amplitude of pixel-to-pixel 
fluctuations is not unusually high
\item ``spectral colours'' have values $|a-b| < 0.1$, $|a-c| < 0.3$ and $|b-c| <
0.3$ ensuring that the spectra do not exhibit significant large scale
gradients.  $a$, $b$ and $c$ are calculated by averaging the flux in three wavelength
regions largely free of OH sky emission (7000:7200\,\AA\, [a], 8100:8250\,\AA\, [b]
and 9100:9180\,\AA\, [c]).
\end{enumerate}

We further only accept spectra with a minimum of 3800 good pixels (``{\small
NGOOD}'' parameter) over the entire wavelength range of the spectrum, to eliminate
spectra with substantial numbers of missing pixels.  Following a trial run of
the PCA, 300 spectra are identified as outliers in the distribution of principal
components, therefore dominating particular components, and discarded (see
Section \ref{sec_pca}).  Application of these selection criteria leaves
\nskyspec sky spectra.  Finally, the spectra were normalised to
have a median flux of zero over the wavelength range $6700-9180\,$\AA.  None of the
results in the paper depend on the details of the criteria used to define the
subset of sky spectra to be used in the PCA-analysis.

\subsubsection{Poisson error normalisation}

The true error on each pixel in each spectrum is made up of components due to
Poisson noise and systematic sky-subtraction errors.  Our method for removing
sky residuals relies on determining the best-fit PCA-components via a minimum
least-squares criterion over all pixels within the $6700-9180\,$\AA\, wavelength
range.  A pixel which varies greatly between spectra will be weighted highly
during the generation of the PCA-components and in the subsequent reconstruction
of the sky-residuals present in individual spectra.  For example, 
there will be a greater variance among spectra at wavelengths in the
vicinity of strong OH emission lines, purely due to Poisson noise and
independent of whether there is a contribution due to sky-subtraction
errors.  It follows that to achieve effective sky-subtraction it is important to
normalise each spectrum by the Poisson noise expected at each pixel.  Failure to
do so results in over-subtraction for certain pixels, with ``corrected'' pixels
apparently exhibiting fluctuations below the Poisson limit.

\begin{figure}
\vspace*{-3mm}
 \begin{minipage}{\textwidth}
   \hspace*{-0.5cm}
   \includegraphics[scale=0.5]{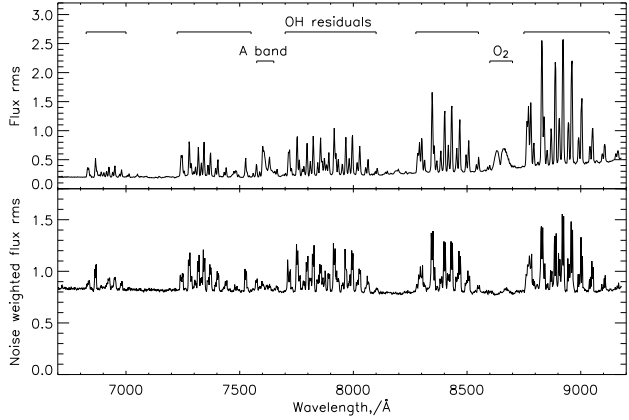}
 \end{minipage}
\vspace*{-2mm}
  \caption{\small Top: the flux rms (67th percentile)
    of the \nskyspec sky spectra.  Bottom: same as top,
    with the flux
    normalised by the associated median SDSS spectroscopic plate noise arrays. 
    The impact of the presence of the OH sky emission lines is evident from
    the increase in the rms (by nearly a factor 2 at maximum),
    even after the empirical scaling of the noise arrays in the SDSS
    spectroscopic reduction pipeline.}
  \label{fig_hedge1}
\end{figure}

Each SDSS spectrum includes a companion error array based on the original
Poisson errors derived from photon counts, CCD read-noise and so forth.
Unfortunately for our application, the error arrays have been systematically
increased in regions of poor sky subtraction.  The procedure is carried out
separately for each plate and details can be found in the data reduction source
code at http://spectro.princeton.edu/idlspec2d\_doc.html\#{\small SKYSUBTRACT}.
Fig.  \ref{fig_hedge1} shows the flux standard deviation\footnote{Unless
otherwise stated estimates of rms amplitudes are taken to be the 67th
percentile of the data.  This is less sensitive to the 
presence of small numbers of extreme, non-Gaussian, outliers.}  (rms) of the
\nskyspec sky spectra with and without normalisation by the
associated SDSS noise arrays.  If the SDSS noise arrays accounted for all the
variance present in the spectra, the lower plot in Fig.  \ref{fig_hedge1} would
be flat.  In fact the SDSS noise arrays account well both for the increased
``continuum'' noise at red wavelengths, the presence of the atmospheric A-band
at $\sim 7600$\AA \ and the contribution from the increased sky emission
associated with the presence of the broad O$_2$ airglow emission centred on
$\sim 8650$\AA.  However, significant additional residual variance resulting
from the incomplete subtraction of the barely resolved OH emission lines
remains.  For many applications the resulting scaled error arrays, which better
reflect the true error in each spectrum, are an improvement.  However, using the
modified SDSS noise arrays for our sky-subtraction procedure does not allow it
to reach its full potential, due to the excessive down-weighting of strong OH
features.

It is not possible to recover directly the original Poisson errors that we
require, so instead we develop an empirical approximation from investigation of
several spectroscopic plates prior to sky subtraction, kindly made available to
us by the SDSS project (E.  Switzer, P.  Macdonald and D.
Schlegel). These data confirmed the expectation that scaling of the
error arrays is applied by the SDSS reduction pipeline predominantly
at positions of OH emission features. 
The same noise scaling is applied to all spectra
on a plate, allowing us to derive a single rescaling for each plate.  For each
one of the 574 plates we calculate the median noise at each wavelength
increment
(pixel) from the SDSS error arrays of the sky spectra.  The
continuum noise is calculated over the full wavelength range of
interest interpolating across pixels containing OH sky emission lines.
The amplitude of
this continuum noise is dominated by counting statistics, and rises towards the
red (Fig. \ref{fig_hedge1}; upper panel) due to an increase in the sky level and instrumental
effects.

\begin{figure}
\vspace*{-3mm}
 \begin{minipage}{\textwidth}
   \hspace*{-0.5cm}
   \includegraphics[scale=0.5]{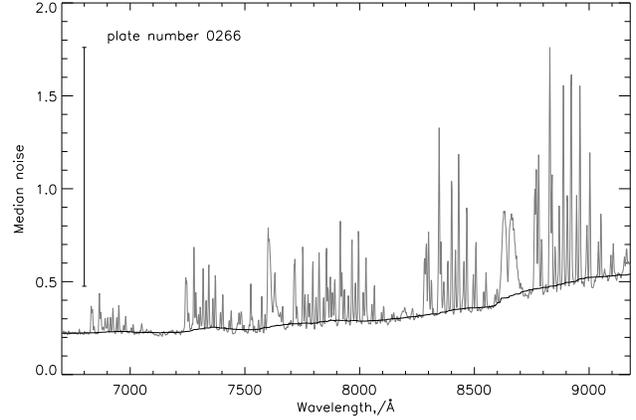}
 \end{minipage}
\vspace*{-2mm}
  \caption{\small The median noise array for plate number
    0266 (grey) and the estimated continuum overplotted (black). The bar to
    the left hand side indicates the amplitude of max$(n-c)$ for this plate,
    which occurs at the wavelength of $8829\,$\AA. } 
  \label{fig_egplate}
\end{figure}

We use a simple function to approximate the rescaling of the noise,
$S_i$, for each pixel $i$ on each plate:
\be \label{eq_scale}
S_i = 1+\left[\frac{(n-c)_i}{{\rm max}(n-c)}\right]^\alpha \times \beta 
\ee
where $n$ is the median noise of the plate, $c$ the continuum noise,
and max$(n-c)$ the maximum noise, above the continuum level, over all
pixels. Fig. \ref{fig_egplate} shows an example plate median noise
array, with estimated continuum and max$(n-c)$ marked. $\alpha$ and $\beta$ are
determined empirically as described below and
represent, respectively, the relative amount of rescaling between
pixels with differing amounts of noise, and the total rescaling of the
pixel with the highest noise above continuum level. As the exact
effects of the SDSS noise scaling are not understood, the form of this
function is chosen to provide flexibility in both the total rescaling
and dependence with wavelength.

\begin{figure}
\vspace*{-3mm}
 \begin{minipage}{\textwidth}
   \hspace*{-0.5cm}
   \includegraphics[scale=0.5]{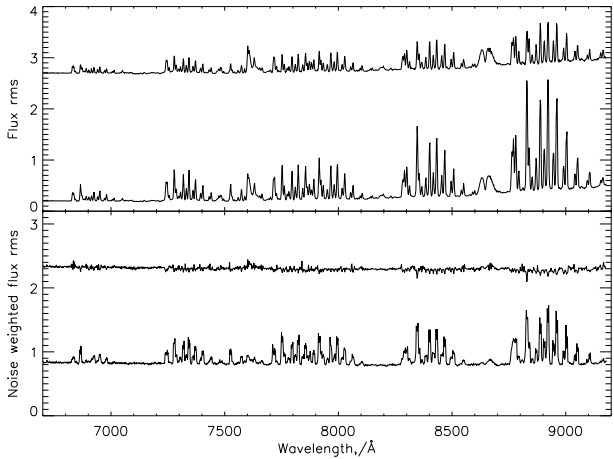}
 \end{minipage}
\vspace*{-2mm}
  \caption{\small Top: the flux rms (67th percentile)
    of the \nskyspec sky spectra. The lower (upper) 
    spectrum is before (after) the
    sky-residual subtraction is performed. Bottom: same as top, with
    flux normalised by the associated median SDSS spectroscopic plate noise
    arrays, scaled according to Eq. \ref{eq_scale} ($\alpha=1,\beta=0.3$). For clarity, the 
    upper spectra
    have been displaced vertically by 2.5 units (upper panel) and 1.5
    units (lower panel).
    The effectiveness of the sky-residual subtraction procedure is seen by the
    reduction by more than half of the systematic spikes in the top
    figure, and complete removal in the bottom figure.} 
  \label{fig_hedge2}
\end{figure}

Anticipating the results of Section \ref{sec_recon}, the effectiveness of the
PCA sky-residual subtraction procedure can be seen in Fig.  \ref{fig_hedge2}.
The top panel shows the effect of the sky-residual subtraction on the raw rms of
the sky spectra:  the lower spectrum shows the
uncorrected rms as a function of wavelength (reproducing the top spectrum
from Fig.  \ref{fig_hedge1});  the upper spectrum shows the raw rms
after application of the
sky-residual subtraction scheme.  Note how the features due to the atmospheric
A-band ($\sim 7600\,$\AA) and O$_2$-emission ($\sim 8450\,$\AA) are
unaffected, while the height of the spikes associated with the OH
emission lines is greatly reduced.  The variation with wavelength of
the rms, post sky-residual subtraction, is explained almost entirely by
counting statistics.  The bottom panel of  Fig.  \ref{fig_hedge2}
shows the rms of the sky spectra weighted by our
modified noise arrays (with $\alpha=1,\beta=0.3$): the lower and upper
spectra show the rms before and after
subtraction of the sky residuals respectively.

\begin{figure}
\vspace*{-3mm}
 \begin{minipage}{\textwidth}
   \hspace*{-0.5cm}
   \includegraphics[scale=0.5]{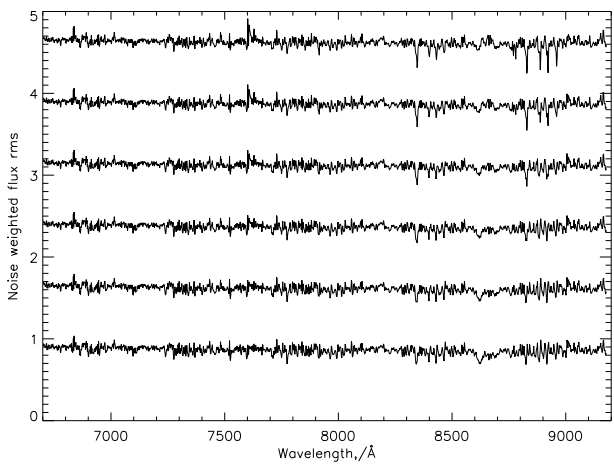}
 \end{minipage}
 \vspace*{-2mm}
  \caption{\small  Flux rms (67th percentile) of the \nskyspec sky
   spectra after sky-residual subtraction is performed, with
    flux normalised by the associated median SDSS spectroscopic plate noise
    arrays, scaled according to Eq. \ref{eq_scale}. From bottom to top
    $\beta = 0,0.1,0.2,0.3,0.4,0.5$ and $\alpha=1$.} 
  \label{fig_beta}
\end{figure}

Fig. \ref{fig_beta} shows the effect of varying the noise rescaling
($\beta$ in Eq. (\ref{eq_scale})) applied to the sky spectra
prior to creating the eigenspectra. From bottom to top
$\beta=0,0.1,0.2,0.3,0.4,0.5$. For low values of $\beta$ (small amounts of
rescaling of the noise arrays) the noise weighted rms is
approximately constant with wavelength, i.e. the sky-residual subtraction
procedure is not removing signal below the Poisson limit for any
pixels, although the procedure may also not be removing the full systematic 
contribution to the errors.
As beta increases, initially the rms remains constant until a point is
reached where significant
dips appear in the red half of the spectra. The presence of these dips
show that an unphysical situation has resulted, where, following the
rescaling, the noise falls below the Poisson limit for those pixels.
The same effect is seen if we undertake the sky residual subtraction
without weighting the flux by the noise arrays, i.e. the noise is
constant, independent of wavelength. The fact that we can
rescale the noise arrays by some amount before these unphysical dips
appear, indicates that the SDSS noise arrays have indeed been scaled
to take account of the presence of a systematic contribution to the
noise at wavelengths where the OH-emission is significant.

Under-subtraction of the sky-residual signal due to over estimation of
the Poisson errors prevents our sky-residual subtraction method from
reaching its full potential.
Over-subtraction of the sky-residual signal causes downward spikes to appear at
the location of the OH emission lines (Fig.  \ref{fig_beta}), as we
begin to erroneously subtract Poisson noise.
This suggests a simple method to estimate the values of 
$\alpha$ and $\beta$ in Eq. \ref{eq_scale}:  we require the weighted
flux rms, post sky-residual
subtraction, to be as constant with wavelength as possible
whilst maximising the noise rescaling.  We take a grid of values for $\alpha$
and $\beta$ and for each combination first run the PCA on the 
\nskyspec sky spectra with noise normalisation scaled according 
to Eq.  (\ref{eq_scale}), then perform with the sky-residual
subtraction with the same noise weighting.  By calculating the standard deviation of
the rms array, post
sky-residual subtraction, for each value of ($\alpha$, $\beta$) we can
derive the best empirical rescaling of the SDSS noise arrays:  we require the
greatest value of $\beta$ that does not introduce upward or downward
spikes, whilst $\alpha$ allows purely for potential wavelength
variations. We
find that $\alpha=1$ and $\beta=0.3$.  Our final results are insensitive to
small changes in $\alpha$ or $\beta$, and the measured value of $\alpha=1$
suggests that the empirical rescaling of the noise arrays within the SDSS
spectroscopic pipeline is predominantly a function of the height of the OH
spikes above the continuum noise.  By systematically decreasing the
amplitude of the noise arrays by 30\% at positions of OH lines whilst
retaining a constant noise weighted
rms with wavelength, we substantially increase the effectiveness of the method,
without over-subtracting residual sky features beyond the limit set by counting
statistics.

We emphasise that it is not possible to derive the true noise array, based on
counting statistics, for each SDSS spectrum from the information contained in
the publically available SDSS data releases.  However, the close to constant rms
as a function of wavelength achieved following our empirical rescaling of the
noise arrays (Fig.  \ref{fig_hedge2}; bottom panel) is strong evidence that,
while approximate, the scheme does achieve a highly effective correction,
reducing the rms close to the level set by counting statistics at all
wavelengths.  If a version of our sky-residual subtraction scheme could be
applied to the SDSS spectra using the true noise arrays, based on counting
statistics, then further improvements should result.

\subsubsection{Identification of sky pixels}

The strong sky emission lines, which lead to the presence of systematic
sky-residuals, occupy only a fraction of the $6700-9180\,$\AA \ wavelength range and
it is necessary to identify those pixels which suffer significantly from sky
subtraction errors.  The affected pixels can be found via the calculation of the
rms of the \nskyspec sky spectra, normalised by the corresponding
SDSS noise arrays (as shown in the lower panel of Fig.  \ref{fig_hedge1}).  All
pixels above a threshold value are defined as ``sky pixels''.  The ``non-sky
pixels'' are not included in the PCA sky-residual removal procedure but provide
an empirical estimate of the level of Poisson noise (i.e. excluding systematic
sky-residuals) present in each spectrum.  Adopting a threshold level of 0.85 for
the sky/non-sky boundary, the scheme results in 670 sky pixels and 697 non-sky
pixels.  The results shown in subsequent sections are not sensitive to
reasonable variations in the threshold level.  Finally, to minimise their effect
during the calculation of the PCA components, sky pixels in each spectrum where
the associated SDSS noise array is zero (i.e. no data) are set to the mean
value of the pixel in all the sky spectra that contain data.

\subsection{Principal component analysis} \label{sec_pca}

\begin{figure*}
\vspace*{-6mm}
 \begin{minipage}{\textwidth}
   \hspace*{-0.5cm}
   \includegraphics[scale=1.]{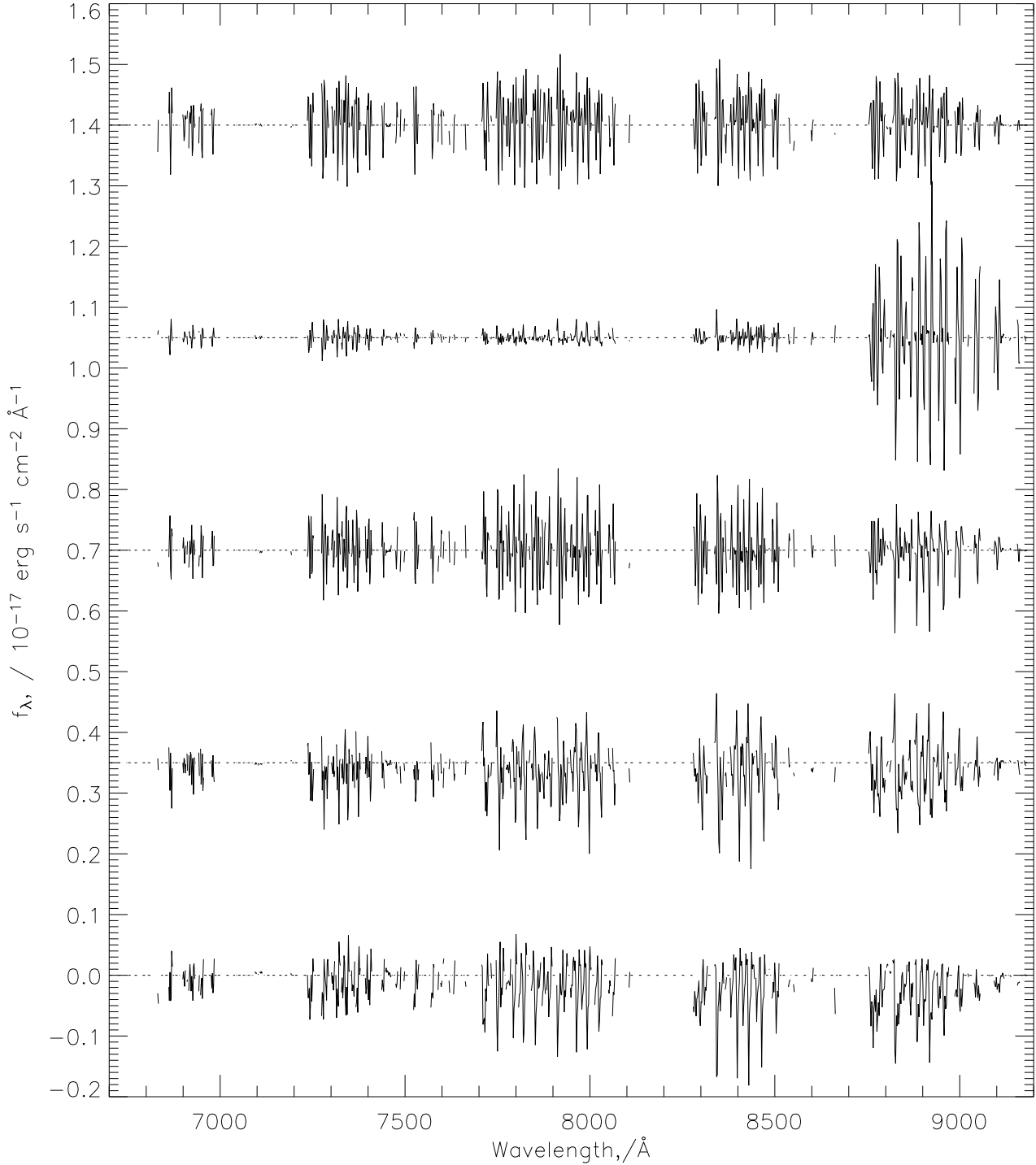}
 \end{minipage}
\vspace*{-5mm}
  \caption{\small From top to bottom, the first 5 sky principal
    components, each offset vertically from the next by a value
    0.35. A horizontal dashed
    line indicates zero flux in each component.}
  \label{fig_temp}
\end{figure*}

PCA is a simple statistical method for data reduction which looks for components
in a dataset that vary the most.  The technique can be visualised by imagining
an $M$-dimensional array, where $M$ is the number of pixels in the spectra.  Each
spectrum is then represented by a point in the $M$-dimensional array; PCA searches
for lines of greatest variance through the array, each one being orthogonal to
all previous lines.  The line of greatest variance defines the first PCA
component, and so on until the line of least variance defines the $M^{th}$ PCA
component.  The projection of an individual spectrum back onto a PCA component
gives the amplitude of the component contained in that particular
spectrum. A spectrum is reconstructed by summing the principal components multiplied by
the relative amplitudes for that spectrum.
Appendix \ref{ap_pca} presents the mathematics of PCA.

Having created a sample of sky spectra containing only those
pixels with sky signal and weighted by our empirically derived noise array for
the relevant plate, we run the PCA.  The output consists of $M$ principal
components which are $M$ pixels long.  Each component has an associated
variance, which gives the
percentage of the total variance of the dataset contained in that component. 

As PCA can be sensitive to spectra with particularly unusual features in which
we are not interested, we remove spectra which dominate the signal in individual
components after one trial run.  The pruning, of 300 spectra in total, is
achieved by removing those spectra with principal component amplitudes more than
$5\sigma$ from the mean.  The PCA is then rerun on the resulting sample of
sky spectra.  Fig.  \ref{fig_temp} shows the first 5 principal
components resulting from the analysis.  Note the distinctive, correlated
features present in each component.

Once the components are created, they can be used as templates to
reconstruct the input sky spectra and later the sky residuals in the
object spectra (Section \ref{sec_reconobj}). This is done by projection of each
spectrum onto the components and summation of the components weighted
by the projection coefficients (Eq. \ref{eq_recon} and
\ref{eq_recon}). The reconstruction may then be subtracted from the
sky spectra leaving residual-free spectra, termed sky-residual
subtracted spectra.

\subsubsection{The number of components}

The number of components to use in the reconstruction is not well defined.  The
use of too many results in the artificial suppression of noise below the Poisson
limit, with the PCA acting as an (undesirable) high-frequency filter.  The use
of too few components means that the removal of sky residuals is sub-optimal
and, in some cases, the overall quality of the spectra can decrease.

\begin{figure}
\vspace*{-3mm}
  \begin{minipage}{\textwidth}
  \hspace*{-0.2cm}
  \includegraphics[scale=0.5]{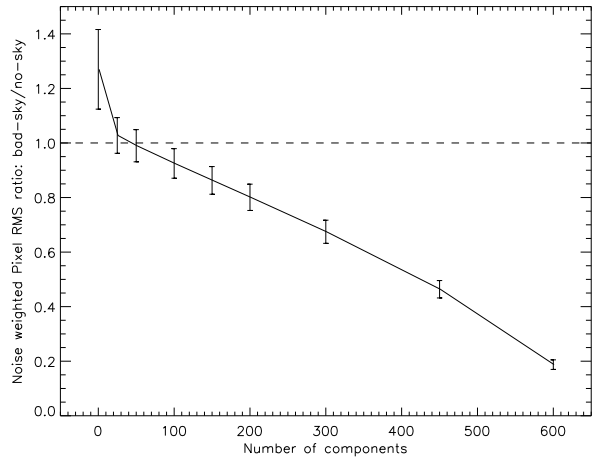}
\end{minipage}
\vspace*{-2mm}
  \caption{\small The mean ratio of sky to non-sky flux rms over all sky
  spectra in our sample as a function of the number of components used
  during sky residual reconstruction. Fluxes are weighted by their
  respective scaled plate noise arrays. The horizontal dashed line
  indicates where the two rms values are equal. The error bars show
  the standard deviation of all objects in the sample. }
  \label{fig_nrecon}

\end{figure}

Fig.  \ref{fig_nrecon} shows the mean ratio of the rms of the
noise-weighted flux in the sky pixels to that in
the non-sky pixels\footnote{Calculation of
the non-sky rms requires a robust estimator unaffected by non-Gaussian outliers,
whereas the sky rms must remain sensitive to outliers.  Therefore, while a 67th
percentile rms is used for the former, the standard deviation of the data is
used for the latter.}, for the \nskyspec sky-residual subtracted sky spectra,
as a function of the number of components used in the reconstructions.  The rms of
the non-sky pixels in each spectrum remains constant and the ratio decreases
monotonically as more components are used in the reconstructions, with an
increasing fraction of the noise in the sky pixels removed.

The reduction of the noise in the sky pixels below the noise in the non-sky
pixels is clearly unphysical, and we therefore estimate the number of components
to employ in the reconstruction of each spectrum by adopting the non-sky pixel
rms of that spectrum as a reference.  The reconstruction of a spectrum proceeds
one component at a time, with the rms ratio calculated for the
sky-residual subtracted spectrum.  Reconstruction is stopped when the
rms ratio reaches unity, i.e. 
when the noise weighted flux rms is the same for the sky and non-sky
pixels. The scheme is largely self-calibrating.  For example, in spectra with high
S/N the systematic sky residuals typically contribute only
marginally to the sky-pixel noise, the rms ratio thus starts close to unity and
only a small number of components are necessary to achieve equality in the rms
ratio. 

\begin{figure}
\vspace*{-3mm}
 \begin{minipage}{\textwidth}
\hspace*{-0.2cm}
  \includegraphics[scale=0.5]{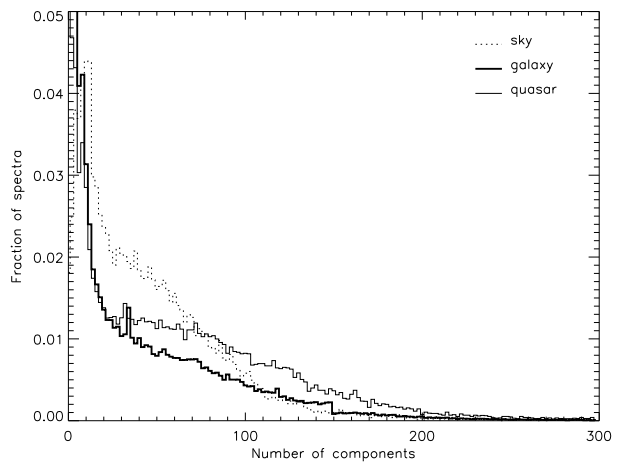}
\end{minipage}
\vspace*{-2mm}
  \caption{\small Histograms showing the number of components used to
    reconstruct the sky signal in 
    the sky (dotted line), galaxy (thin line) and quasar (thick line) DR2
    samples. The y axis is truncated in order to show the majority of
    the objects, 18\% (5\%) of galaxies (quasars) require zero components. } 
  \label{fig_hist}
\end{figure}

The large number of pixels that contribute, combined with the robust estimator
of the non-sky rms, mean the amplitude of the non-sky rms is well determined,
providing an excellent indicator of when to halt the
reconstructions. Fig. \ref{fig_hist} shows the distribution of the
number of components used for DR2 sky, galaxy and quasar spectra. For
bright objects the contribution to the noise from imperfections in the 
sky subtraction is small and few components are required to bring the
sky and non-sky rms to the same level.
The presence of many spectra with high S/N produces the
``spike'' at low
component numbers in both the galaxy and quasar sample histograms. The
fainter limiting magnitude of the quasar 
sample make sky-residuals a problem in a larger fraction of the
spectra and, on average, more components are required to remove the 
systematic errors present following the default SDSS-pipeline sky-subtraction. 
To limit the effect of very occasional poor estimation of the non-sky
rms we set the maximum number of components to 150 and 200 for the
galaxy and quasar samples respectively.

\subsection{Reconstructing sky residuals in an object spectrum}\label{sec_reconobj}

In the preceding sections we have developed a procedure that achieves a
substantial reduction in the amplitude of the systematic sky residuals in
spectra that possess no large scale signal, i.e.  the successfully
sky spectra are known, by definition, to possess zero signal at all wavelengths.
We now turn to the more interesting application of removing the
systematic sky-residuals from the SDSS science spectra of targets such
as galaxies and quasars.

The problem is made
tractable by the success of the sky-subtraction procedure performed as part of
the standard SDSS spectroscopic pipeline.  Detailed examination of the
properties of the sky spectra shows that both the mean level and
the large scale shape of the sky spectrum to be subtracted from each spectrum
have been determined to very high accuracy.  As a result, there is effectively
no sky ``continuum'' to remove, rather the problem is confined to removing the
high-frequency structure due to the presence of the strong OH sky emission
lines. The key goal is to ensure that object continuum and intrinsic absorption and
emission line features are not removed as a result of the procedure used to
reduce the amplitude of the systematic sky-residuals. 

As it is not necessary to identify any sky continuum present, we can remove the
continuum of the object spectrum using a median filter before projection of the
spectrum onto the principal components derived from the sky
spectra.  The smallest filter size can be found from median filtering the sky
spectra as it is important that the filter is unaffected by the OH
residuals (about 40 pixels). As the filter size is increased above this minimum, some
features intrinsic to the object fail to be removed. This generally causes an increase
in the rms of the non-sky pixels used as a reference point to halt the
reconstruction (see previous Section) and therfore a decrease in the
number of components used and slight decrease in the effectiveness of
the sky-residual subtraction procedure. However, the final effect is
small and insignificant for a reasonable range of filter sizes (up to
about 80 pixels). We have used a filter size of 55 pixels throughout
this paper. 
In certain circumstances, it may be desirable to reduce the filter scale in order
to follow the broad emission line profiles in quasars closer to the
line centroids. 

Narrow line features present the greatest challenge, as these can be mistaken
for an OH sky residual by the PCA.  The advantage of PCA in this task is that it
looks for patterns in the spectrum, linking lines together which are correlated
in the input sky spectra, and weights all bins equally.  However,
if an emission or absorption feature occurs exactly at the location of an OH
line, some combination of principal components can sometimes be found to
reconstruct the feature, without increasing the rms in the rest of the spectrum
significantly.  Such behaviour is particularly likely if the line feature lies
in a noisy part of the spectrum.  For most applications in which the
sky-residual subtraction scheme might be employed, it is possible to mask known
emission and absorption features, thereby circumventing the problem.  In Section
\ref{sec_cat} we show how such masked features are unaffected by the
sky-residual subtraction procedure.  The disadvantage is that real sky features
which happen to fall in the masked regions do not contribute to the projection
onto the principal components and the sky-residual subtraction may not be fully
effective in these regions.  In some potential applications, it is not
known in advance where absorption and emission features may occur and we present
such an example in Section \ref{sec_dla}.  Features are masked by removing the
relevant pixels during projection onto the principal components.  Alternatively,
replacing the pixels with the local mean results in almost identical
reconstruction. We similarly mask bad pixels (where the SDSS noise array is
set to zero).

\begin{figure*}
\vspace*{-5mm}
 \begin{minipage}{\textwidth}
   \hspace*{-0.5cm}
   \includegraphics[scale=0.86]{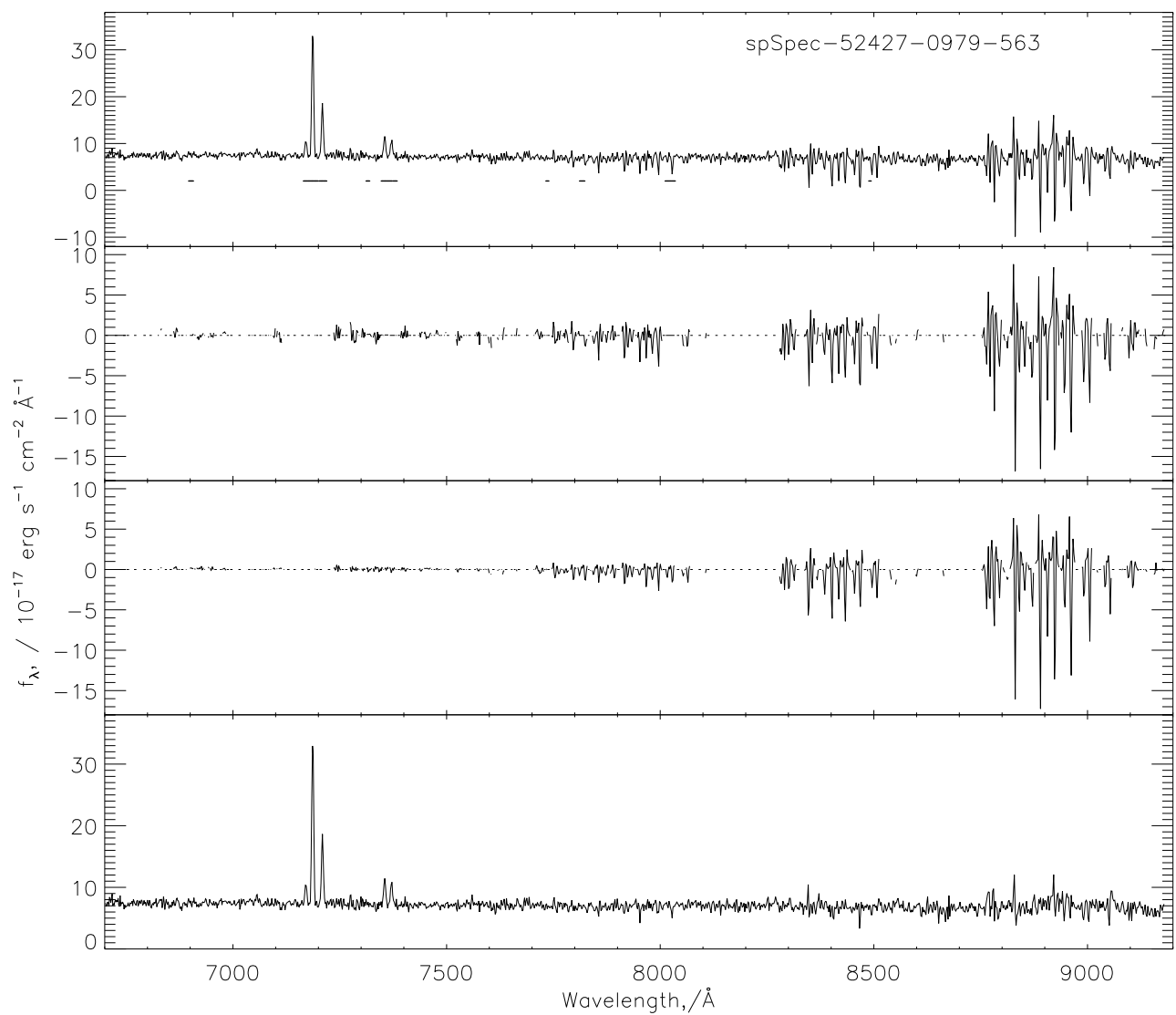}
 \end{minipage}
\vspace*{-4mm}
  \caption{\small 1 (top): The wavelength region $6700-9180\,$\AA, of the SDSS
    galaxy spectrum spSpec-52427-0979-563. Pixels included in the feature mask
    are indicated by black horizontal bars under the spectrum; 2:
    The same spectrum but showing only ``sky'' pixels, regions between
    sky pixels are joined by a dotted line; 3: The
    reconstructed sky spectrum; 4 (bottom): The final sky-residual subtracted
    galaxy spectrum.}
  \label{fig_gal1}
\end{figure*}

Once the object spectrum has been continuum subtracted, we divide by the SDSS
noise spectrum of the relevant plate and project onto the sky principal
components (Eq. \ref{eq_pcs}), leaving out those pixels identified as
possibly containing emission and absorption features.  The resulting principal
component amplitudes are used to reconstruct the residual sky-subtraction signal
(Eq. \ref{eq_recon}) and the reconstruction is subtracted from the object
spectrum, including masked pixels.  Re-multiplication by the noise spectrum,
followed by the addition of the object continuum, returns the object spectrum
cleaned of OH sky emission line residuals.  Fig.  \ref{fig_gal1} illustrates the
process of sky-residual subtraction on a typical SDSS galaxy spectrum.

\section{Application of method to SDSS spectra}\label{sec_recon}

Each class of spectroscopic science targets has different spectral
characteristics.  The effective application of the sky residual subtraction
scheme requires the removal of the large scale ``continuum'' signal from the
target object and the identification of wavelength intervals where narrow
emission or absorption features may be present.  The extremely high
identification success rate achieved in the SDSS means that the wavelengths of
strong absorption and emission features in the spectra of stars and galaxies are
known; there are only 3,002 spectra without secure identifications among the
329,382 object spectra included in DR2.  Potential systematic biases in the sky
residual subtraction can be prevented by masking the wavelength intervals that
include such features.

The exact nature of the pre-processing applied to spectra prior to the
implementation of the sky residual subtraction depends on the scientific goal.
However, in subsequent sections we describe schemes that are likely to have wide
application in the analysis of the 3 main classes of SDSS science targets:
galaxies, quasars and stars.

\begin{figure*}
 \begin{minipage}{\textwidth}
\hspace*{-1cm}
   \includegraphics[scale=1.05]{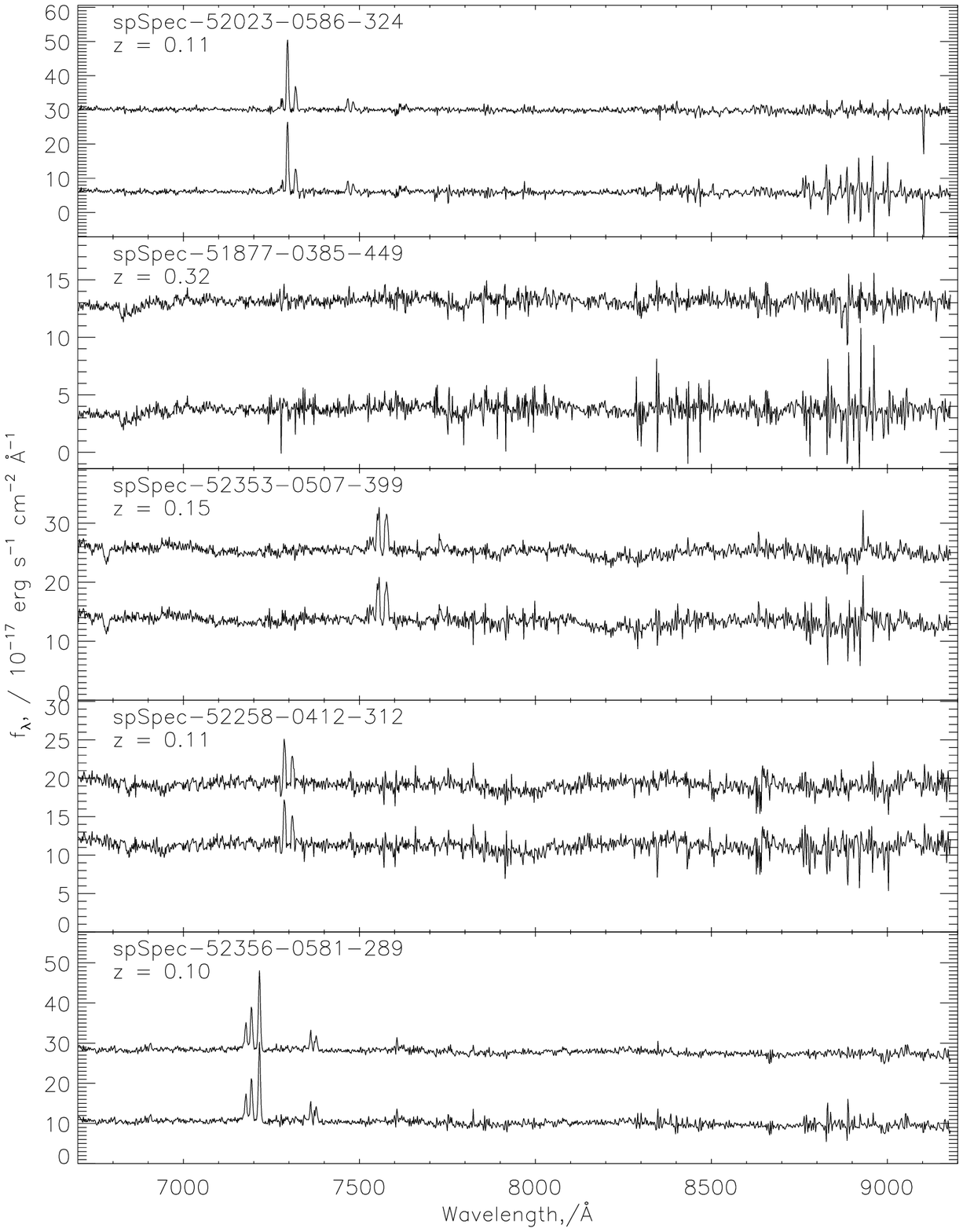}
 \end{minipage}
  \caption{\small Examples of sky-residual subtraction applied to
    galaxy spectra. In each panel the lower spectrum is the raw SDSS
    data and the upper spectrum is after application of the sky-residual
    subtraction procedure. The upper sky-residual subtracted spectrum is
    offset for clarity.}
  \label{fig_gal2}
\end{figure*}

\begin{figure*}
 \begin{minipage}{\textwidth}
\hspace*{-1cm}
   \includegraphics[scale=1.05]{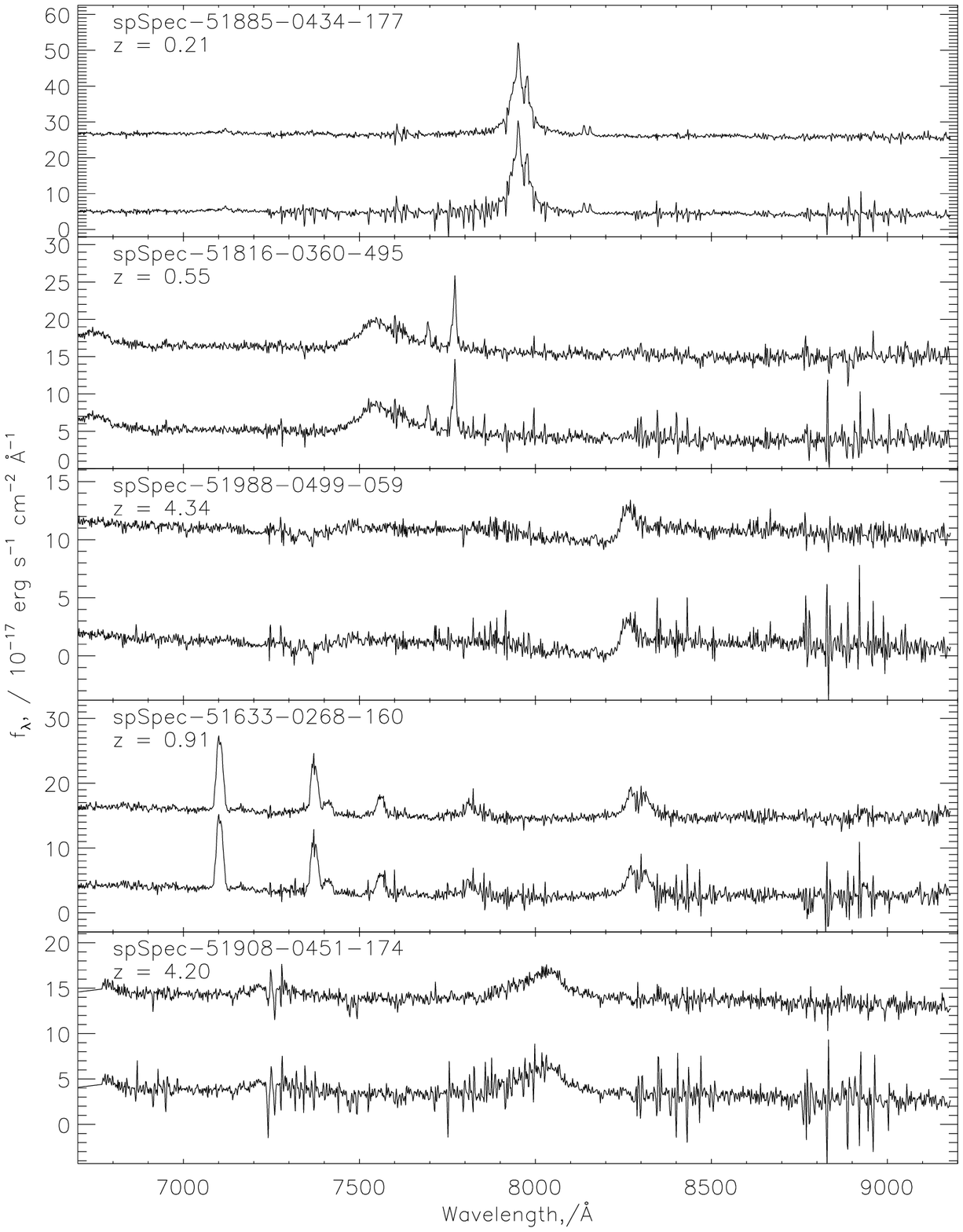}
 \end{minipage}
  \caption{\small Examples of sky-residual subtraction applied to
    quasar spectra. In each panel the lower spectrum is the raw SDSS
    data and the upper spectrum is after application of the sky-residual
    subtraction procedure.  The upper sky-residual subtracted spectrum is
    offset for clarity.}
  \label{fig_qso}
\end{figure*}

\subsection{Galaxies}

The DR2 catalogue contains 249,678 unique objects spectroscopically classified
as galaxies.  The strong systematic trends in the absorption- and emission-line
properties of the galaxies as one moves from early through to late type objects
necessitates a sub-classification of the population prior to the application of
the sky-residual subtraction.  Therefore, we classify each galaxy as either an
absorption line, emission line or extreme emission line object.  The spectral
type parameter (``{\small ECLASS}'') in the SDSS catalogue provides the basis
for the absorption line ({\small ECLASS}$<-0.05$) and emission line ({\small
ECLASS} $\ge -0.05$) classification.  The extreme emission line galaxies are
identified through the presence of emission lines with very large equivalent
widths (EW) (${\rm EW_{H\alpha}}>200$\AA, $z<0.4$; or ${\rm EW_{[OIII]}}>
200$\AA, $z<0.84$).  An appropriate feature mask is then applied depending on
the galaxy type (e.g. see Fig.  \ref{fig_gal1}).  A total of 331/286/469\AA \
are masked in absorption/emission/extreme emission line objects respectively,
over the entire rest-frame wavelength range of $3830-9100\,$\AA, i.e.  in the
observed-frame wavelength range of $6700-9180\,$\AA \ only a small
fraction of pixels are affected by the line masks. Fig. \ref{fig_gal2}
shows examples of galaxy spectra before and after sky residual removal.

\subsection{Quasars}\label{sec_qso}

The DR2 catalogue contains 34,674 unique objects spectroscopically classified as
quasars or high-redshift quasars.  A feature mask includes narrow emission lines
(e.g.  [OIII] $4961, 5008\,$\AA) and $70\,$\AA \
intervals\footnote{Corresponding to approximately half the filter size
in this wavelength range} (observed frame) centred on the
broad emission lines (e.g.  CIV $1550\,$\AA).  The wings of the broad emission features
can, in practice, be regarded as ``continuum'' in the context of the sky
residual subtraction as they are removed by the median filtering.
Fig. \ref{fig_qso} shows examples of quasar spectra before and after sky subtraction.

\subsection{Stars}
There are 42,027 unique objects classified as stars or late-type stars
in the DR2. Due to their single redshift, application of the sky-subtraction
procedure is even simpler. Depending on the final science
required and the type of stars involved, a suitable feature mask and
continuum estimation can be straightforwardly derived but our adopted
median filter scale of 71 pixels, employed for the galaxies and quasars,
produces very satisfactory results.

\section{Tests of method on SDSS science objects}\label{sec_tests}
In this Section we present the quantitative results of applying the
sky-residual subtraction to a variety of object spectra.

\subsection{Galaxy absorption features: The CaII triplet}\label{sec_cat}
\begin{table*}
  \centering
  \caption{\label{tab_catrip} \small Mean EWs, EW-ratios and
    sample variances of the final two CaII triplet lines (8544\AA \
    [2], 8664\AA \ [3]),
    before and after the sky-residual subtraction. $m$ is the width of
    the feature mask applied during reconstruction (see text).}
\vspace{0.2cm}
  \begin{tabular}{ccccccc} \hline\hline
     && \multicolumn{2}{c}{Before subtraction} &
    \multicolumn{2}{c}{After subtraction}\\ \hline
    $m$ & mean(EW2+EW3) & mean(EW2/EW3) & var(EW2/EW3) & mean(EW2+EW3) & mean(EW2/EW3) & var(EW2/EW3)  \\ \hline
   1 & -4.86 & 1.46 & 0.35 & -4.86 & 1.46 & 0.32\\
   2 & -4.85 & 1.46 & 0.35 & -4.86 & 1.46 & 0.33\\
   3 & -4.85 & 1.46 & 0.35 & -4.86 & 1.46 & 0.32\\
  \end{tabular}
\end{table*}

\begin{figure*}
 \begin{minipage}{\textwidth}
\hspace*{-1cm}
   \includegraphics[scale=1.05]{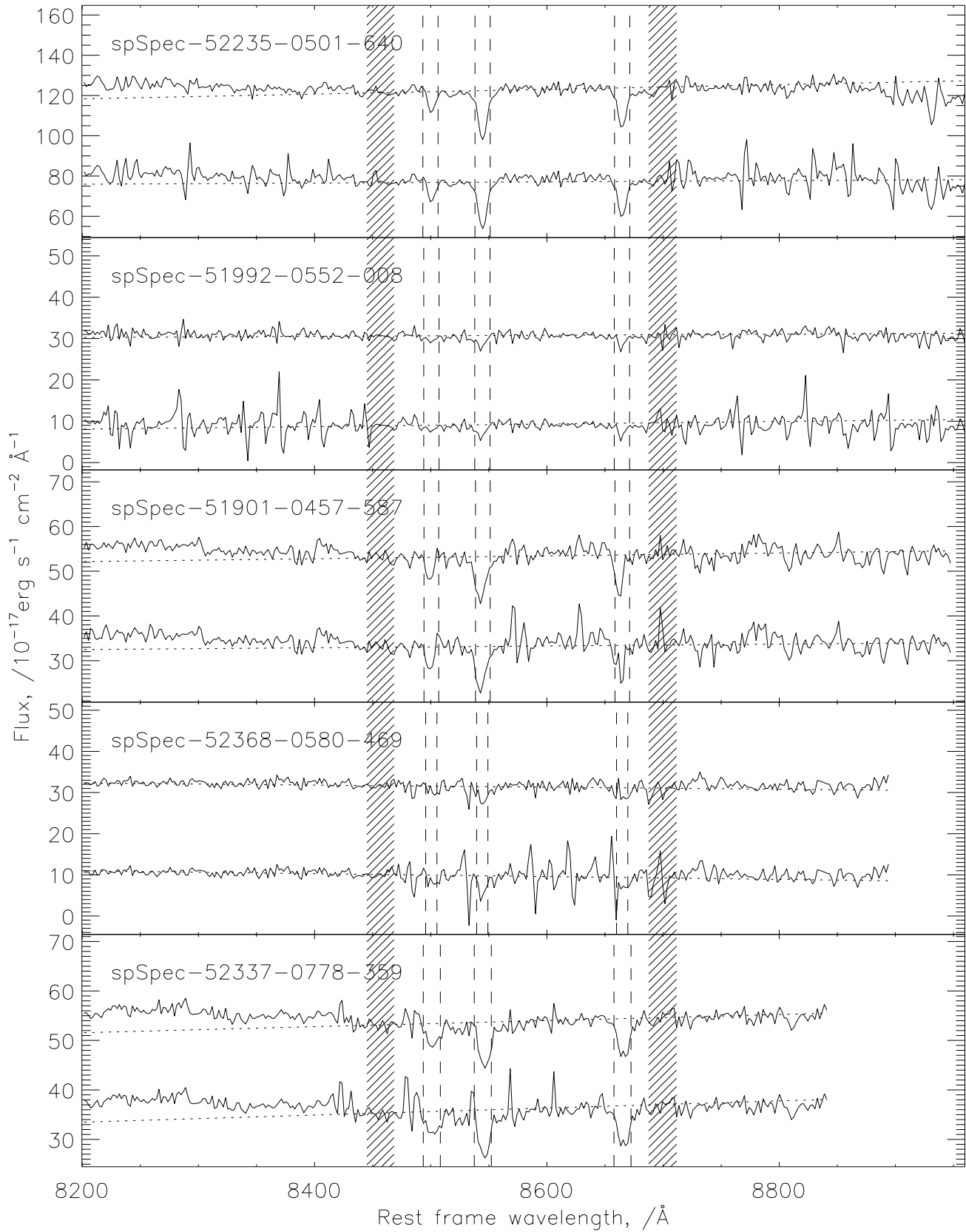}
 \end{minipage}
  \caption{\small Example SDSS spectra in the region of the CaII
    triplet, before (lower) and after (upper) sky-residual subtraction. Vertical
    lines indicate where equivalent widths of lines
    are measured, shaded regions are where the continuum is
    estimated. The inferred continuum is
    overplotted (dashed line). In each
    case the upper corrected spectrum is offset for clarity.}
  \label{fig_catrip}
\end{figure*}

\begin{figure}
 \begin{minipage}{\textwidth}
\hspace*{-0.5cm}
   \includegraphics[scale=0.5]{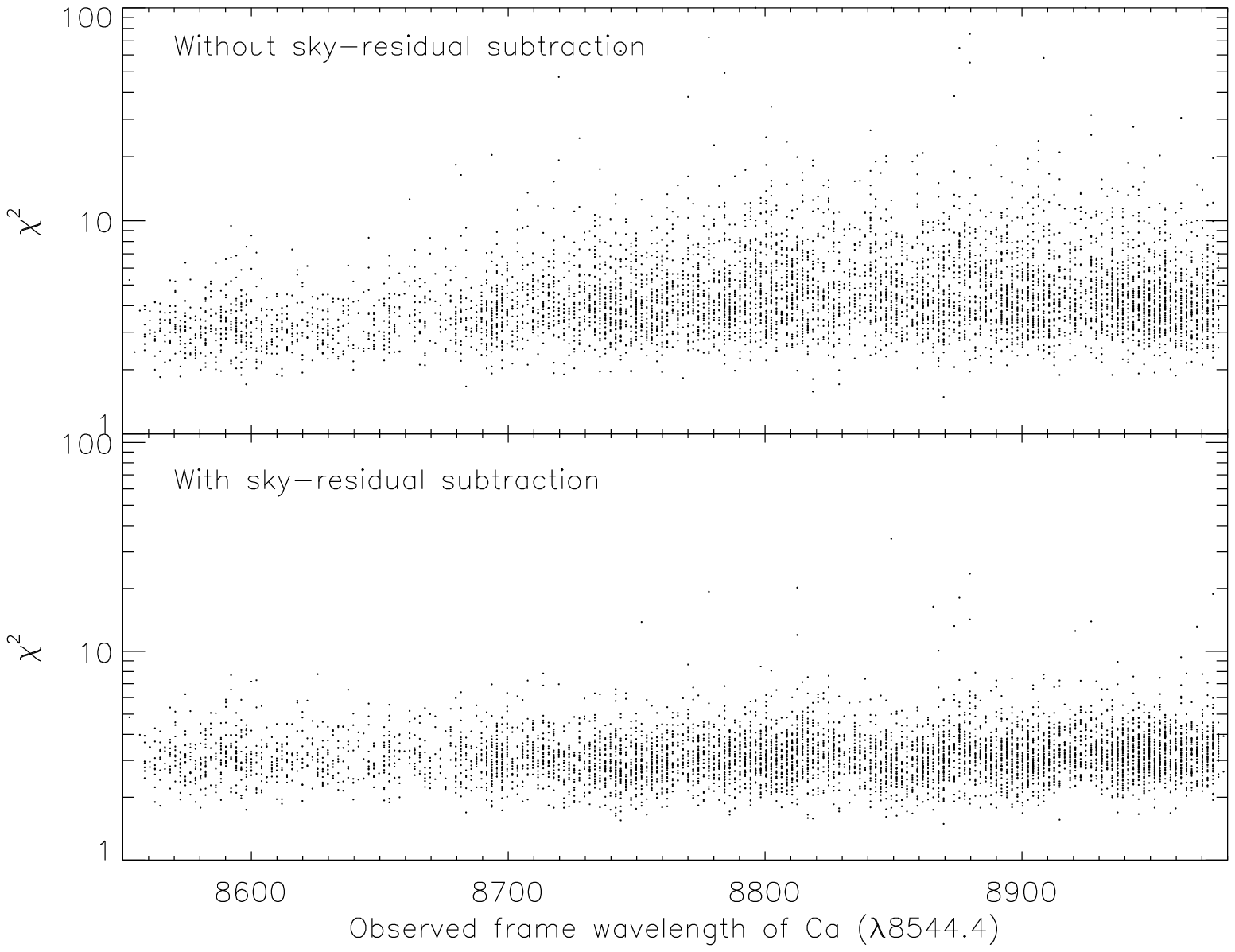}
 \end{minipage}
  \caption{\small Resulting reduced-$\chi^2$ between a scaled matched-filter
     and $\sim 6000$ spectra with detected CaII triplet, as a function of
     the observed wavelength of the 8544.4\AA \ line. Upper (lower) panel
     is for spectra without (with) sky-residual subtraction. All galaxies
     in SDSS
     DR3 with S/N in the R band of $<$ 15 and redshifts $z < 0.05$ 
     are searched for the CaII
     triplet at the redshift of the galaxy. The sky-residual
     subtracted spectra are treated identically to the original
     dataset. The samples plotted result from adopting a $3\sigma$ detection
     threshold for the presence of the CaII triplet.}
  \label{fig_chi2}
\end{figure}

The far red spectrum of galaxies is where old stellar populations emit most of
their light.  Contamination by light from hot, young stars and any blue
``featureless'' AGN continuum is reduced compared to shorter wavelengths.  The
impact of Galactic reddening is also much reduced \citep{1997PASP..109..883R}.
The CaII triplet (8500.4, 8544.4, $8664.5\,$\AA), the strongest absorption
feature in this region, is visible in stellar spectra of all but the hottest
spectral types.  The CaII triplet EW is considered a good estimator of
luminosity class in high metallicity objects and a useful metallicity indicator
in metal-poor systems \citep[e.g.][]{1989MNRAS.239..325D}.  The relatively
narrow intrinsic widths of the lines ($\sigma\approx50\rm{km\,s}^{-1}$) and the
increased velocity resolution per \AA, twice that at blue optical wavelengths,
makes the CaII triplet ideal for studying the internal kinematics of galaxies
\citep{1984ApJ...286...97D,1990MNRAS.242..271T}.  In spectra of intermediate
S/N, the principle limitation to the use of the CaII triplet is the presence of
a large number of strong OH emission lines.  The SDSS spectra allow the
measurement of the the CaII triplet in galaxies out to redshifts $z\lsim0.06$.

The CaII triplet is ideally placed to investigate any potential bias in line
feature properties caused by the sky-residual subtraction procedure.  A large
sample of galaxies with significant CaII triplet absorption and measured
velocity dispersions from the SDSS spectroscopic data reduction pipeline is
readily identified. Furthermore, performing our own line search on low
S/N SDSS spectra allows an immediate demonstration of the potential
benefits afforded by the sky-residual subtraction technique.

\subsubsection{Equivalent width line ratios}
Galaxy velocity dispersion ($\sigma_v$) is only measured in the
standard SDSS data
reduction pipeline for objects with {\small ECLASS} $<-0.02$, $z<0.4$ and {\small
SPECCLASS} = ``{\small GALAXY}'' \footnote{see
http://cas.sdss.org/astro/en/help/docs/algorithm.asp}.  We select a sample of
7760 galaxies according to the criteria:  $z < 0.054$, to provide
enough spectrum for continuum estimation; $70<\sigma_v<
420\,\rm{km\,s}^{-1}$; $\rm{S/N}>10$ in the r-band as recommended on the SDSS website;
CaII[8544.4] and CaII[8664.5] absorption lines detected in the SDSS catalogue at
$>4\sigma$ significance.  We remove from this sample those galaxies which are
untouched by the sky subtraction procedure (their non-sky rms is equal to or
greater than their sky rms).  As we have selected high S/N objects through the
requirement of significant CaII triplet line detection, this removes a further
 $\sim 1000$ galaxies, leaving a sample of around 5500 (dependent on the width of
the mask used for the CaII triplet lines).

The centre of each line is given by the line wavelength in the SDSS catalogue;
the feature mask width and the region over which the line EW is measured are set
as $\pm m \sigma_v$ and $\pm n \sigma_v$ respectively.  We use a value of $n=2$
which generally spans the widths of the lines well, $m$ can take different
values to investigate any bias caused by the sky-residual subtraction procedure.
The continuum level at each wavelength in the region of the absorption lines is
determined using a linear interpolation between the median flux contained in two
bands, (8444.3:8469.3\AA\, and 8687.4:8712.4\AA).  Fig. \ref{fig_catrip} shows
examples of SDSS galaxy spectra in the CaII triplet region before and after
sky-residual subtraction.  The regions defining the absorption lines and
continuum bands are indicated.

The rest-frame EW is calculated as 
\be 
{\rm EW} = \sum_{i=1}^{N}\Delta(W_i)(1-\frac{f_i}{c_i}) 
\ee 
where N is the total number of
pixels, $\Delta(W_i)$ the width of pixel $i$ in \AA \ (rest-frame), $f_i$ the
observed flux, and $c_i$ the continuum.

The total EW is defined as the sum of the two strongest absorption lines at
$8544\,$\AA \ and $8664\,$\AA.  Inclusion of the other, weaker, line often increases the
noise \citep[e.g.][]{1989MNRAS.239..325D}.  While the total EW varies with
galaxy type, the ratio of the line EWs remains constant.  Table \ref{tab_catrip}
presents the mean total EW, and mean and variance of the ratio of first to
second EWs for the galaxy sample, before and after sky residual subtraction, and
for mask widths of $m=1$, 2 and 3.  Provided the feature mask is broad enough to
include the wings of the absorption lines, the mean total EWs are not
significantly different.  The mean EW ratios and variances also remain constant even with the
narrowest feature mask.  

On average, the sky-residual subtraction produces improvements in the S/N of
spectra within the regions of the masked absorption features; the feature mask
excludes the absorption line wavelengths from contributing to the PCA
reconstruction of the sky residual but the sky-residual subtraction is applied
to all wavelengths.  The key result of the test using the properties of the CaII
triplet is that, providing the features are masked prior to the PCA
reconstruction of the sky-residual signal, the properties of the features are
not systematically biased.

\subsubsection{Improvement in feature detection quality}
A practical illustration of the improvement afforded by the
sky-residual subtraction procedure is seen in the results of a
conventional matched-filter search \citep{1985MNRAS.213..971H} for the
presence of the CaII triplet in relatively low S/N spectra. The DR3
release, used in this case simply to increase the sample size,
contains $\sim 15000$ galaxy spectra with S/N in the R band of $<$ 15.  The original and sky-residual subtracted spectra were treated identically. Adopting a nominal $3\sigma$ threshold for detection results in 5702 and 5662 detections respectively, 5119 of which are common to both. Close to the $3\sigma$ S/N threshold, twice as many spurious detections of the CaII triplet  "in emission" exist in the original sample, providing evidence that the $\sim10\%$ of lines unique to each sample are statistically more reliable in the sky-subtracted data set. However, the most significant difference apparent in the properties of the detected features is in the quality of the fit between the matched-filter template and the data. For each CaII triplet detection a goodness-of-fit is calculated based on the sum of the squared deviations between the data and the scaled template. In Figure \ref{fig_chi2} this is plotted versus the observed frame wavelength of the 8544\AA \ line. The overall scatter in the $\chi^2$ distribution at wavelengths where OH lines are present ($>$8650\AA) is significantly reduced in the case of the sky-residual subtracted spectra (the median $\chi^2$ for detections with wavelength $>$8650\AA \ decreases by 27\%).  The systematic trend as a function of wavelength is also essentially removed. A significant fraction of the original spectra present extremely poor fits where the observed frame wavelength of the triplet coincides with a particularly strong OH line. At wavelengths $>$8650\AA, the fraction of detections with $\chi^2$ values exceeding twice the median value decreases from 8\% to just 1\% in the case of the sky-residual subtracted galaxies. 

\subsection{Absorption features in damped Ly$\alpha$ systems}\label{sec_dla}

\begin{figure*}
 \begin{minipage}{\textwidth}
\hspace*{-0.5cm}
   \includegraphics[scale=1.]{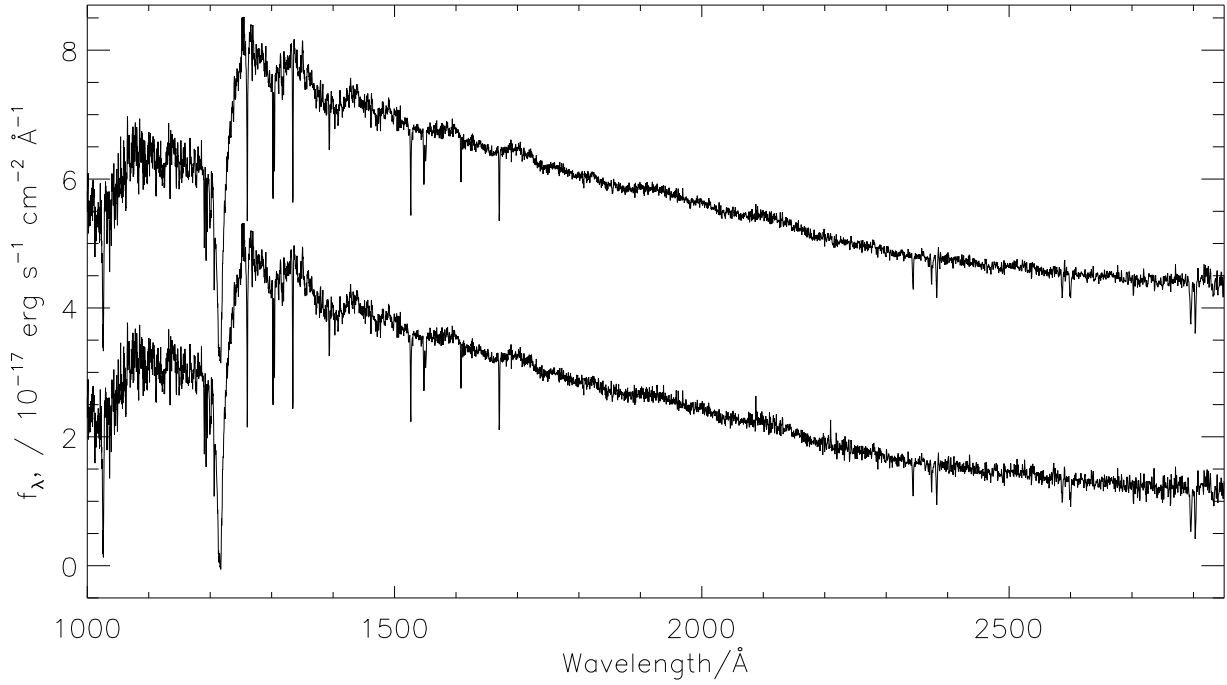}
 \end{minipage}
  \caption{\small Composite spectrum of 68 DLAs identified in the
    SDSS-DR1 by \citet{astro-ph/0403391}, before (lower) and after
    (upper) sky-residual subtraction. The absorption features were masked prior to
    sky-residual subtraction. No attempt was made to remove the
    underlying composite quasar spectrum.}
  \label{fig_dla1}
\end{figure*}

\begin{figure*}
 \begin{minipage}{\textwidth}
\hspace*{-0.5cm}
  \includegraphics[scale=1.]{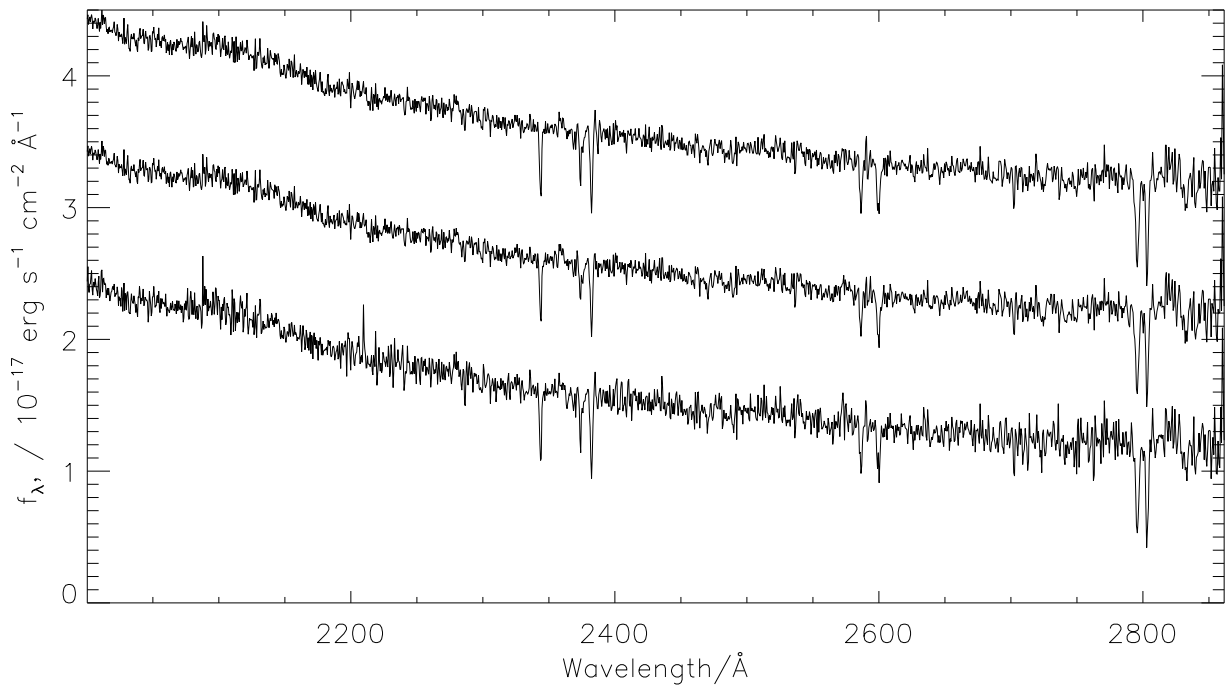}
 \end{minipage}
  \caption{\small Composite spectra of 63 DLA systems 
showing the rest-frame
wavelength range $2000-2850\,$\AA \ that includes strong absorption
features due to Mg and Fe. The 3 spectra show the composite
before sky-residual subtraction (lower), after sky-residual subtraction with
absorption features unmasked (middle), and with absorption features masked
(upper).}
  \label{fig_dla2}
\end{figure*}

\begin{table*}
  \centering
  \caption{\label{tab_dla} \small Equivalent widths of absorption
    features in a composite of 68 SDSS-DR1 DLAs before and after sky residual
    subtraction: (a) absorption features unmasked; (b) $14\,$\AA\,
    around 
absorption features masked. Wavelengths of continuum and measurement
    regions are given, together with the number of spectra where the
    wavelength of the line falls between 6700 and $9180\,$\AA, therefore
    contributing to the composite. }
\vspace{0.2cm}
  \begin{tabular}{cccccccc} \hline\hline

    Line ID & N & cont1 & cont2 & line & before & $\rm{after^ a}$ & $\rm{after^b}$  \\ \hline
  FeII[2344.9] & 47 & 2327,2337 & 2353,2363 & 2340.9,2347.9 & -0.65 & -0.58 & -0.66 \\
FeII[2375.2] & 46 & 2357,2367 & 2392,2402 & 2371.2,2378.7 & -0.50 & -0.44 & -0.54 \\
FeII[2383.5] & 45 & 2357,2367 & 2392,2402 & 2379.0,2386.5 & -0.68 & -0.62 & -0.66 \\
MgII[2796.4] & 11 & 2760,2790 & 2810,2825 & 2792.0,2799.5 & -1.89 & -1.71 & -1.81 \\
MgII[2803.5] & 10 & 2760,2790 & 2810,2825 & 2799.5,2807.0 & -1.75 & -1.55 & -1.73 \\

  \end{tabular}
\end{table*}

The detection of intervening absorption features in quasar spectra is an example
of an investigation in which it is not possible to simply mask the wavelength
regions containing features before the application of the sky residual
subtraction.  That is not to say that the sky-residual subtraction scheme is not
of potential interest.  The improvement in the quality of the spectra is such
that the S/N of specific features can increase, albeit at the potential cost of
modification of the feature properties.  In practice, a hybrid scheme, involving
one or more iterations, with detected features masked in a second application of
the sky-residual subtraction, is straightforward to implement.

In any particular application, a full understanding of the advantages of
employing the sky-residual subtraction scheme would require the use of
(straightforward) Monte Carlo simulations to quantify the probability of feature
detection as a function of wavelength.  Such an investigation is beyond the
scope of the present paper but to illustrate the impact of the sky-residual
subtraction on unmasked absorption features we create a composite spectrum of
damped Ly$\alpha$ (DLA) systems identified by \citet{astro-ph/0403391}.  Each
quasar spectrum is shifted to the rest frame of the DLA absorber and normalised
to possess the same signal in the absorber rest-frame wavelength interval
$1250-1800\,$\AA.  The composite spectrum is then constructed using an
arithmetic mean of all the spectra with signal at each wavelength, taking care
to account for the flux/\AA \ term in the SDSS spectra.  Fig. \ref{fig_dla1}
shows the resulting composite spectrum, calculated with and without the
application of the sky-residual subtraction to the constituent quasar spectra.
As in Section \ref{sec_qso} the centres of prominent broad emission lines in the
quasars are masked during the sky-residual subtraction.

Fig. \ref{fig_dla2} shows 3 versions of the composite spectrum, focusing on
the rest-frame wavelength region that derives from observed-frame wavelengths
$\lambda > 6700\,$\AA.  The 3 versions of the composite were calculated using
quasar spectra without sky-residual subtraction (bottom), with sky-residual
subtraction (middle) and with sky-residual subtraction {\it and} the absorption
features masked.  The absorption-line mask consisted of 14$\,$\AA \ intervals,
in the absorber rest-frame, centered on each absorption line.

Absorption-line equivalent widths (EWs) are measured in the absorber rest-frame,
using the method of Section \ref{sec_cat}.  Table \ref{tab_dla} includes the EWs
of the absorption features in the 3 composite spectra, together with the
wavelength intervals used in the line and continuum measurement.  For each
composite the same continuum is used for measuring EWs, in this case calculated
from the spectrum with lines masked, although in practice any
consistent continuum would suffice.  The sky-residual subtracted composite, with
masking of absorption lines, provides the best reference.  As expected, the EWs
of several of the absorption lines in the unmasked composite are systematically
reduced.  However, the illustration is a ``worst case'' example of the
sky-residual subtraction scheme involving only a single iteration.  A second
application of the sky-residual subtraction using a mask based on the
identification of features following the first application, offers the prospect
of significant improvement in the identification and measurement of features at
initially unknown wavelengths.



\subsection{Composite quasar spectra}\label{sec_compqso}

\begin{figure*}
 \begin{minipage}{\textwidth}
   \includegraphics[scale=0.8]{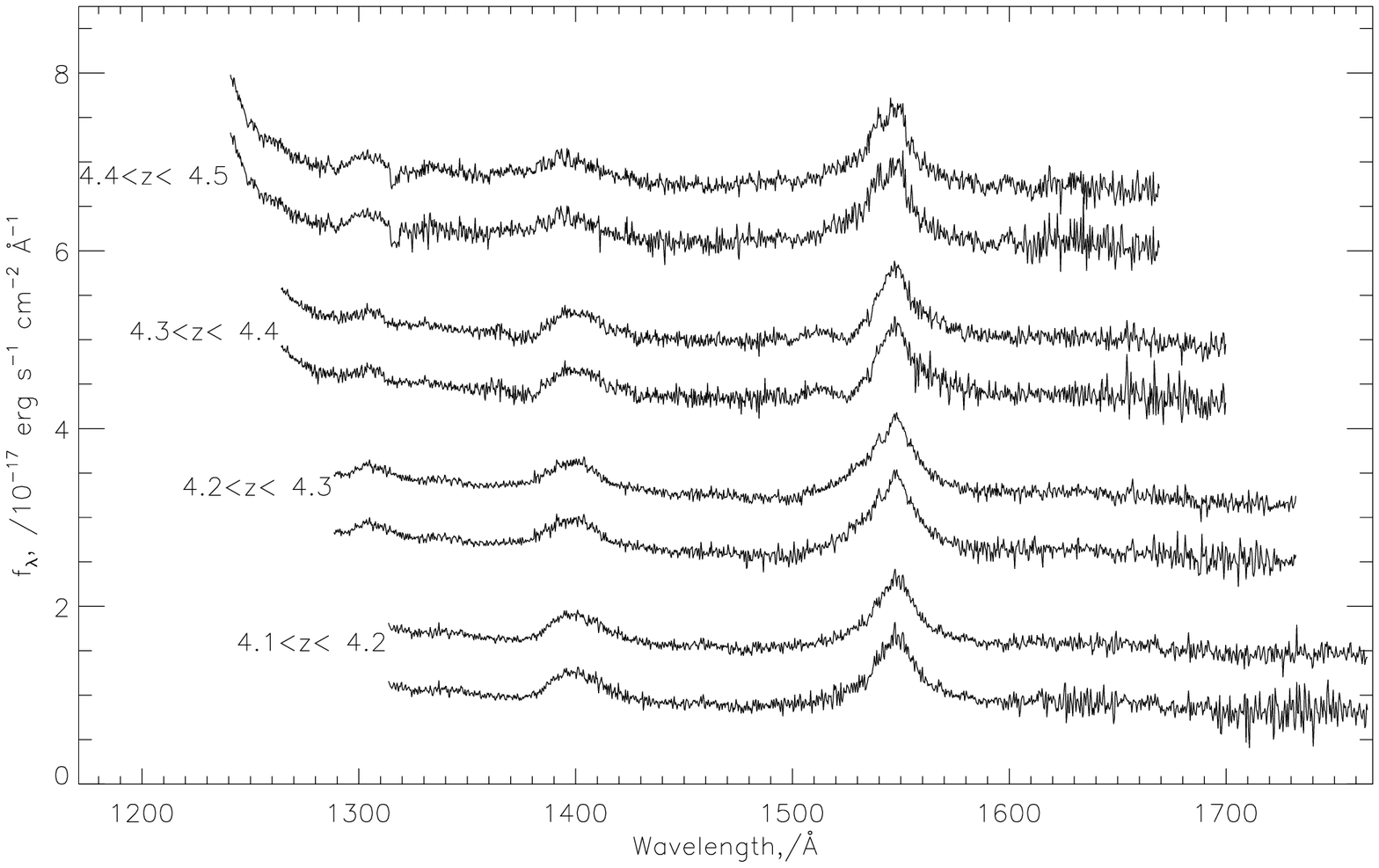}
 \end{minipage}
  \caption{\small Composite spectra of $z\sim 4$ quasars including the
SiIV+OIV $\lambda 1400$ and CIV $\lambda$1549 emission lines. The
observed-frame spectra fall at wavelengths $\lambda > 6700\,$\AA.
For each pair of spectra, the lower composite is created
directly from the SDSS spectra, and the upper composite from the
same spectra after subtracting sky residuals. From bottom to top
the composites consist of 41, 35, 24 and 20 spectra and the
improvement in error weighted absolute deviation from the underlying
spectrum (see text) is 23\%, 17\%, 17\% and 18\%.}
  \label{fig_qsocomp}
\end{figure*}

Poor sky subtraction affects most those spectra with low S/N, and the fainter
quasars, including those at high redshift ($z > 4$), suffer from significant
contamination from sky subtraction residuals.  The extended wavelength range of
the SDSS spectra combined with a redshift range for the quasars of $z \simeq
0-5$, allows the construction of composite spectra over unprecedented ranges in
rest-frame wavelength and luminosity \citep{2004apsd.conf...21V}.  However, the
number of quasars that occupy the extremes of the wavelength and luminosity
ranges is relatively small and the sky subtraction residuals at $\lambda >
6700\,$\AA \ limit both the determination of the continuum properties and the
detection of weak emission features.  We create composite spectra of all quasars
in the SDSS with $z>4.1$, in redshift slices of $\Delta z = 0.1$.  Each input
quasar spectrum is normalised to have a mean flux of 1 between $1250-1600\,$\AA,
prior to combination using an arithmetic mean.  Fig.  \ref{fig_qsocomp}
illustrates how the sky-residual subtraction technique improves the quality of
these composites.  Subtracting the underlying spectrum using a median filter and
removing the centers of emission lines from the calculation, we calculate the
error weighted absolute deviation from zero on the composites with and without
sky-residual subtraction.  The average improvement in S/N over the entire
wavelength region due to the sky subtraction is about 20\% for all redshift bins.

\section{Summary}

The SDSS spectra represent a vast improvement in data quantity, quality and
wavelength coverage over previous surveys.  The sky-subtraction carried out by
the SDSS reduction pipeline has in general achieved excellent results, however
the common problem of the subtraction of undersampled OH emission lines results
in substantial systematic residuals over almost half the spectral range.  A
technique is presented to remove these residual OH features, based on a
principal component analysis of the observed-frame sky spectra which have been
sky subtracted along with all other spectra in the SDSS data releases.  The
scheme takes advantage of the high degree of correlation between the residuals
in order to produce corrected spectra whose noise properties are very close to
the limit set by counting statistics.

The sky-residual subtraction scheme is applied to the SDSS DR2 catalogue and is
immediately applicable to the recent SDSS DR3 release.  The precise form of the
implementation depends on the spectral energy distributions of the target
objects and the scientific goal of any analysis.  As a consequence, the main
classes of target objects, galaxies, quasars and stars, are treated slightly
differently.  Constructing composites of high-redshift quasars before and after
sky-residual subtraction illustrates the significant increase in the overall S/N
of the spectra that is achieved.  The application of the procedure when narrow
absorption or emission features are present requires the use of a mask to
prevent the sky-subtraction scheme suffering from bias.  Both the
calcium-triplet in low redshift galaxies and metal absorption lines in damped
Ly$\alpha$ systems at high redshifts fall in the wavelength range affected by
the OH-forest ($6700-9200\,$\AA).  By employing these two rather different
examples it is shown that masking absorption and emission features prior to the
reconstruction of the sky-residual signal in each spectrum ensures that the
properties of the features themselves are not systematically biased in the
resulting sky-residual subtracted spectra.

The significant improvement in S/N achieved over some $2000\,$\AA \ of a large
fraction of SDSS spectra, particularly for the fainter objects such as the
high-redshift quasars, should benefit a wide range of scientific investigations.
We have made available datafiles, IDL code and a comprehensive guide on using
the code, with which the procedure can be applied to all SDSS spectra
({\tt http://www.ast.cam.ac.uk/research/downloads\\/code/vw/}).  Further
improvements should be possible if the sky-residual subtraction procedure is
adapted to run on the SDSS spectra using the original noise arrays based on
counting statistics alone.  The nature of the sky-residual subtraction scheme is
such that application to other large samples of spectra should be relatively
straightforward.

\section*{acknowledgements}
Many thanks to Eric Switzer, David Schlegel and Patrick McDonald of
the SDSS team who made available information and answered questions
about the SDSS spectroscopic reduction pipeline. We would also like to
thank the referee, Simon Morris, for his careful
reading of the manuscript and helpful comments and suggestions. VW
acknowledges the award of a PPARC research studentship.

Funding for the Sloan Digital Sky Survey (SDSS) has been provided by
the Alfred P. Sloan Foundation, the Participating Institutions, the
National Aeronautics and Space Administration, the National Science
Foundation, the U.S. Department of Energy, the Japanese
Monbukagakusho, and the Max Planck Society. The SDSS Web site is
http://www.sdss.org/.

The SDSS is managed by the Astrophysical Research Consortium (ARC) for
the Participating Institutions. The Participating Institutions are The
University of Chicago, Fermilab, the Institute for Advanced Study, the
Japan Participation Group, The Johns Hopkins University, Los Alamos
National Laboratory, the Max-Planck-Institute for Astronomy (MPIA),
the Max-Planck-Institute for Astrophysics (MPA), New Mexico State
University, University of Pittsburgh, Princeton University, the United
States Naval Observatory, and the University of Washington.

\bibliographystyle{mn2e}

\begin{appendix}
\section{Principal Component Analysis}\label{ap_pca}
For completeness we include the basic mathematics of PCA. Further
details can be found in \citet{1975kendall} and, for
astronomical applications, in \citet{1984MNRAS.206..453E} and \citet{1987mda..book.....M}.
For an $N$ spectra by $M$ pixel data array $\{X_{ij}\}$, where $1\le i\le
N$, $j\ge 1$ and $k\le M$, the covariance matrix is 
\be
C_{jk} = \frac{1}{N}\sum_{i=1}^N \rm{X}_{ij}\rm{X}_{ik}.
\ee
The eigenvectors ($\ul{e}$, principal components in the language of this
paper) and eigenvalues ($\lambda$) of the covariance matrix are 
\be 
\ul{C} \ul{e}_j = \lambda_j \ul{e}_j.
\ee
It can be shown that $\ul{e}_1$ is the axis along which the variance
is maximal, $\ul{e}_2$ is the axis with second greatest variance, and
so on until $\ul{e}_M$ has the least variance. The principal
component amplitudes for each noise weighted input spectrum $\ul{f}$ are given by
\be \label{eq_pcs}
a_j = \ul{f} \cdot \ul{e}_j.
\ee

The $M$ eigenvectors can be used as a basis set on which to project any
spectrum of the same dimensions. The principal component amplitudes of this
projection and the eigenvectors can then reconstruct the input spectrum
from the templates: 
\be \label{eq_recon}
f_k = \sum_{j=1}^M a_j e_{jk}.
\ee
The reconstructed spectrum only contains information present in the
templates, which may or may not be a fair representation of the input
spectrum. In general as the variance of the eigenvectors decrease, so does the
useful information contained in the spectra, allowing PCA to be a
useful form of data compression.  The exact number of eigenvectors
required to reconstruct the input spectrum satisfactorily depends on
the dataset and purpose of the analysis.
\end{appendix}

\end{document}